\documentclass[3p]{elsarticle}
\pdfoutput=1 
\usepackage{graphicx}
\usepackage{amsfonts}
\usepackage{amsmath}
\usepackage{url}
\usepackage{epstopdf}
\usepackage{color}
\usepackage{bm}
\usepackage{multirow}
\usepackage{epsfig}

\DeclareMathAlphabet\mathbit
    \encodingdefault\rmdefault\bfdefault\itdefault
\DeclareOldFontCommand{\bi}{\normalfont\bfseries\itshape}{\mathbit}

\newcommand{\be}{\begin{equation}}
\newcommand{\ee}{\end{equation}}

\def\fakebold#1{\relax\ifvmode\leavevmode\fi%
\ifmmode%
\setbox0=\hbox{$#1$}%
\else%
\setbox0=\hbox{#1}%
\fi%
\kern-.02em\copy0 \kern-\wd0%
\kern .04em\copy0 \kern-\wd0%
\kern-.0125em\raise.02em\box0%
}%

\def\oe{\overline}
\def\beq{\begin{eqnarray}}
\def\eeq{\end{eqnarray}}

\def\dc{\partial}
\def\bc{\begin{center}}
\def\ec{\end{center}}
\def\bt{\begin{tabular}}
\def\et{\end{tabular}}
\def\wi{\widetilde}

\begin{document}

\begin{frontmatter}

\title{Resonant signal reversal in a waveguide connected to a resonator}

\author{Andrey L. Delitsyn}
  \ead{delitsyn@mail.ru}

\address{Higher School of Modern Mathematics MIPT,  1 Klimentovskiy per., Moscow, Russia}

\author{Irina K. Troshina}
 \ead{itroshina@mail.ru}
\address{National Research University Higher School of Economics, Myasnitskaya Street 20, Moscow 101000, Russia}

\begin{keyword}
wave propagation; resonance scattering; tunneling effect; barriers
\end{keyword}

\begin{abstract}
It has been proven that when connecting two infinite semi-cylinders or waveguides with a finite cylinder or resonator at a certain frequency, it is possible to transmit a signal almost completely from one semi-cylinder to another. In this case, the reflected field is arbitrarily small. A very simple technique based on the expansion of the solution in a Fourier series in cylinders and matching the series for the signal and its derivatives in the conjugation boundaries of cylinders of different radii is used for the proof. The main feature of this method is its elementary nature, which allows for a certain class of boundaries to establish resonant scattering effects.
\end{abstract}

\end{frontmatter}

\section{Introduction}

In this paper, resonant wave scattering in a waveguide with a resonator formed by the presence of barriers in the waveguide is considered. The connection of the waveguide with resonators allows, in particular, to turn the wave in the opposite direction after passing through the resonator. Or in a similar way it is possible to shift the wave in the knee-shaped connection. Resonant scattering in waveguides connected to resonators is well known, starting with the works of Rayleigh \cite{1}, and is widely used in technical devices, both for resonant damping and for resonant signal transmission. A large number of works of a physical nature are devoted to this phenomenon. Starting with the works of Arsenyev \cite{2}-\cite{5} this area attracts the attention of mathematicians. Among the works of a mathematical nature we note \cite{6}-\cite{8}. At the same time, these works are quite technical, use the technique of splicing asymptotic expansions and require certain efforts for understanding by specialists in the field of physics and mechanics.
An explanation of the principle of the resonant nature of wave transmission through barriers was given in the work \cite{9}. We use an exceptionally simple technique that does not require knowledge of complex mathematical methods and allows us to clarify the situation with resonant scattering based on elementary methods. All that is required is to apply the expansion of the solution in a Fourier series and elementary methods of the theory of Hilbert spaces. At the same time, the applied method allows us to obtain a number of new results. Despite the fact that in this paper we consider this method as applied to problems in domains that are a union of cylinders, this limitation is not fundamental and the method can be applied to a significantly wider class of domains. Consideration of the problem in the specified particular class of domains corresponds to a large number of problems arising in applications and greatly simplifies the technical application of the method. We consider a problem in a domain Q that is a union of three cylinders - two semi-infinite cylinders $ Q_1, Q_2 $ and a finite cylinder $ Q_3 $ (see Fig. 1). $$ Q_1 = \left\{(x, y) \in \Omega, z \in (-\infty, 0) \right\} . $$
The cylinder $ Q_2 $ is obtained by mirror-symmetrical reflection of the cylinder $ Q_1 $ relative to the plane $ y = 0 $.
$$ Q_3 = \left\{(x, y) \in \Omega_3, z \in (0, a) \right\} . $$

Cylinders $ Q_1 $ and $ Q_2 $ are geometrically identical and are connected to cylinder $ Q $ through small holes. The region is mirror symmetric with respect to the plane $ y = 0 $. In this paper, it will be proved that there is a frequency at which a wave incident on the resonator along one of the cylinders almost completely transforms into an outgoing wave in the second cylinder. In this case, the reflection coefficient of the incident wave in the first cylinder is arbitrarily small.

\begin{figure}[ht]
\centering
\includegraphics[width=105mm,height=75mm]{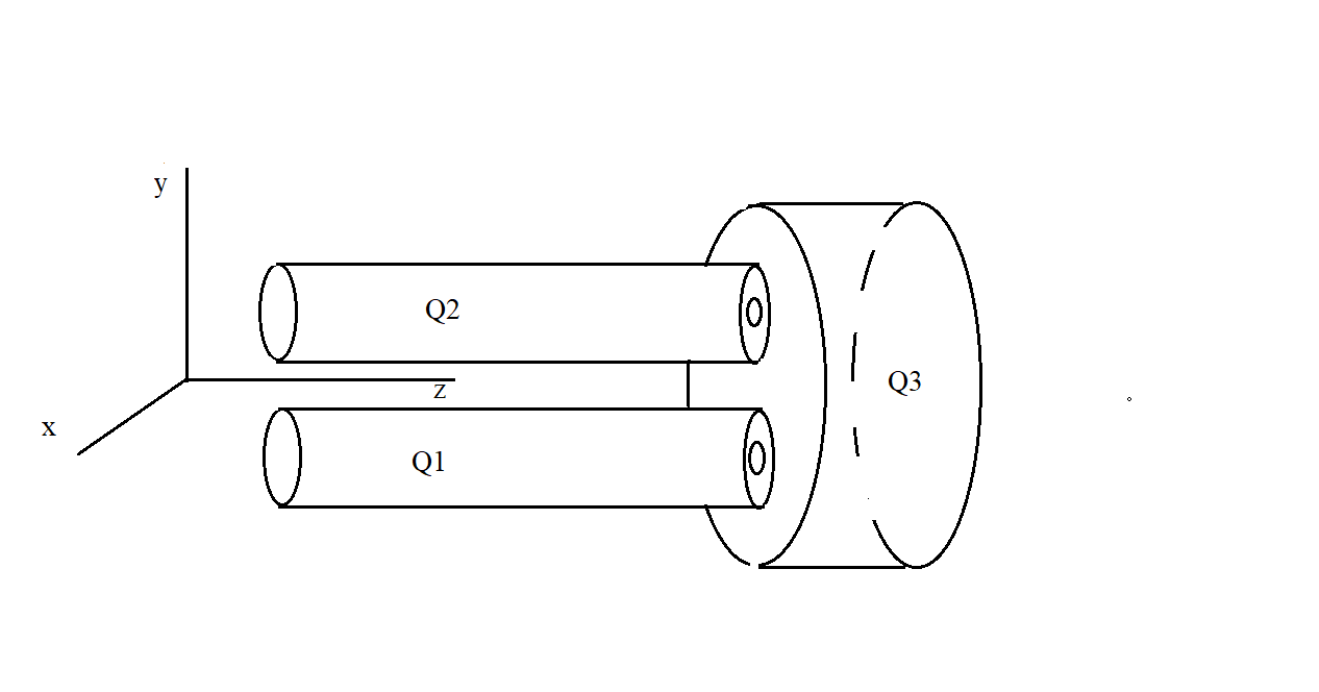}
\caption{Waveguides connected by resonator}
\label{fig2}
\end{figure}

\section{Mathematical formulation of the problem}

The problem is formulated mathematically as follows. The solution to the problem must satisfy the Helmholtz equation in the region $ Q $
\be \Delta u + k^2 u = 0 \ee
with the boundary condition on the boundary of the region $ Q $
\be u|_{\partial Q} = 0 \ee
and the radiation conditions in the cylinder $ Q_1 $ and the cylinder $ Q_2 $.
\be u = e^{i \gamma_1 z} \psi_1(x,y) + r_1 e^{-i \gamma_1 z} \psi_1(x,y)
+ \sum \limits_{n=2}^{\infty} r_n e^{\gamma_n z} \psi_n(x,y) \quad (z < 0) \ee
\be u = t_1 e^{-i \gamma_1 z} \psi_1(x, - y) + \sum \limits_{n=2}^{\infty} t_n e^{\gamma_n z} \psi_n(x, - y) \qquad (z < 0), \ee
where $ \lambda_n, \psi_n $ are the eigenvalues and eigenfunctions of the problem in the cross section $ \Omega $ of the cylinder $ Q $:
$$ - \Delta \psi_n = \lambda_n \psi_n, \quad \quad \psi_n|_{\dc \Omega} = 0 $$
$$ \gamma_1 = \sqrt{k^2 - \lambda_1} , \quad \gamma_n = \sqrt{\lambda_n - k^2} \qquad (n \geq 2) $$
We consider the problem in the range of frequencies $ k \in (\sqrt{\lambda_1}, \sqrt{\lambda_2}) $.
The coefficients $ r_n, t_n $ are unknown and are determined by solving the problem. The coefficient $ r_1 $ is called the reflection coefficient, $ t_1 $ is the transmission coefficient.

We reduce the problem in the domain $ Q $ to a problem in the subdomain $ Q_3 $, which is a subdomain $ Q $ bounded by the plane $ S = 0 $.
Let $ S $ be a section of the region $ Q $ by the plane $ y = 0 $.
Let us consider the problem in a region $ \oe Q $ consisting of a cylinder $ Q_1 $ and $ {\wi Q}_3 $ which is half of the cylinder $ Q_3 $, cut off by the plane $ y = 0 $ for $ y < 0 $. On the section $ S $ we consider two problems, the first with the Dirichlet condition on $ S $
\be \Delta u^D + k^2 u^D = 0, \label{1} \ee
\be u^D|_{\dc Q} = 0, \ee
\be u|^D_S = 0, \ee
\be u^D = e^{i \gamma_1 z} \psi_1(x,y) + r_1^D e^{-i \gamma_1 z} \psi_1(x,y)
+ \sum \limits_{n=2}^{\infty} r_n^D e^{\gamma_n z} \psi_n(x,y) \quad (z < 0), \label{2} \ee
the second with the Neumann condition
\be \Delta u^N + k^2 u^N = 0, \ee
\be u^N|_{\dc Q} = 0, \ee
\be \frac{\dc u}{\dc z}|_S = 0, \ee
\be u^N = e^{i \gamma_1 z} \psi_1(x,y) + r_1^N e^{-i \gamma_1 z} \psi_1(x,y)
+ \sum \limits_{n=2}^{\infty} r_n^N e^{\gamma_n z} \psi_n(x,y) \quad (z < 0), \ee
Continuing the solution of the Dirichlet and Neumann problems in an odd and even way to the domain $ Q $ corresponding to $ y > 0 $
$$ u^D(x, y, z) = -u^D(x, - y, z), $$
$$ u^N(x, y, z) = u^N(x, - y, z). $$
The extended solutions satisfy all boundary conditions in $ Q $ and the same radiation conditions in $ Q_2 $ as in $ Q_1 $.
The half-sum of solutions to the Dirichlet and Neumann problems satisfies the Helmholtz equation in $ Q $, the boundary conditions on the boundary of $ Q $, and the radiation conditions in $ Q_2 $
$$ \frac{1}{2} (u^D + u^N) = \frac{1}{2}(r_1^N - r_1^D) e^{-i \gamma_1 z} \psi_1(x, -y) + \frac{1}{2}
\sum \limits_{n=2}^{\infty} (r_n^N - r_n^D) e^{\gamma_n z} \psi_n(x, -y). $$
Thus, the solution of the scattering problem is reduced to two problems - the Dirichlet problem and the Neumann problem.
Since the reflection coefficient is equal to
$$ 1 + r_1 = (u, \psi_1)_{L_2(\Omega)}, \qquad r_n = (u, \psi_n)_{L_2(\Omega)}, $$
then, having proved that the reflection coefficient for a certain for the Dirichlet problem is equal to $ 1 $, we obtain that the reflection coefficient of the desired problem is arbitrarily close to $ 0 $, provided that the diameter of the hole connecting the waveguide with the resonator is sufficiently small.
Thus, the original problem is reduced to problem (\ref{1})-(\ref{2})

\section{Main equation and the method of its consideration}

As a result, we consider the problem of scattering in a cylinder connected by a small hole to a cylindrical resonator in the region $ \oe Q $.
We represent the solution in the region $ \oe Q $ in the form (\ref{2}).
The coefficients $ r_n $ are expressed in terms of $ u $ as follows
$$ r_1 = (u, \psi_1)_{L_2(\Omega)} - 1, $$
$$ r_n = (u, \psi_n)_{L_2(\Omega)}. $$
Differentiating the solution $ u $ with respect to $ z $ and passing to the limit as $ z \to 0-0 $, we obtain
$$ \frac{\dc u}{\dc z} = 2 i \gamma_1 \psi_1 - i \gamma_1 (u, \psi_1)_{L_2(D)}\psi_1+ \sum \limits_n \gamma_n (u, \psi_n)_{L_2(D)} \psi_n. $$
In the domain $ {\wi Q}_3 $ we represent the solution as
$$ u = c_1 \sin(\beta_1 (z - a)) \chi_1(x, y) + \sum \limits_n c_n \sinh(\beta_n (z - a)) \chi_n(x, y) $$
where $ \mu_n, \chi_n $ are the eigenvalues and eigenfunctions of the problem in the cross section $ {\wi \Omega}_3 $  of the cylinder $ {\wi Q}_3 $:
$$ - \Delta \chi_n = \mu_n \chi_n, \quad \quad \chi_n|_{{\wi \Omega}_3} = 0 $$
$$ \beta_1 = \sqrt{k^2 - \mu_1} , \quad \beta_n = \sqrt{\mu_n - k^2} \qquad (n \geq 2) $$
Differentiating with respect to $ z $ and passing to the limit at $ z \to 0+0 $, we obtain
$$ \frac{\dc u}{\dc z} = -\beta_1 \cot(\beta_1 a) (u, \chi_1)_{L_2(D)}\chi_1
- \sum \limits_n \beta_n \coth(\beta_n a) (u, \chi_n)_{L_2(D)} \chi_n $$
Equating in the hole $ D $ the left and right limit values of the derivative $ \frac{\dc u}{\dc z} $ we get
\be 2 i \gamma_1 \psi_1 - i \gamma_1 (u, \psi_1)_{L_2(D)}\psi_1 + \sum \limits_n \gamma_n (u, \psi_n)_{L_2(D)} \psi_n = \label{15} \ee
$$ -\beta_1 \cot(\beta_1 a) (u, \chi_1)_{L_2(D)} \chi_1 - \sum \limits_n \beta_n \coth(\beta_n a) (u, \chi_n)_{L_2(D)} \chi_n. $$
Thus we arrive at the equation
\be - i \gamma_1 (u, \psi_1)_{L_2(D)} \psi_1 -\beta_1 \cot(\beta_1 a) (u, \chi_1)_{L_2(D)}\chi_1+
\sum \limits_n \gamma_n (u, \psi_n)_{L_2(D)} \psi_n
+ \ee
$$ \sum \limits_n \beta_n \coth(\beta_n a) (u, \chi_n)_{L_2(D)} \chi_n = -2 i \gamma_1 \psi_1. \label{Eq} $$
Let us rewrite the equation (\ref{15}) in the form
\be (- i \gamma_1 - 1) (u, \psi_1)_{L_2(D)} \psi_1 - \beta_1 \cot(\beta_1 a) (u, \chi_1)_{L_2(D)} \chi_1 +
\gamma_1 (u, \psi_1)_{L_2(D)} \psi_1 + \ee
$$ \sum \limits_n \gamma_n (u, \psi_n)_{L_2(D)} \psi_n
+ \sum \limits_n \beta_n \coth(\beta_n a) (u, \chi_n)_{L_2(D)} \chi_n = -2 i \gamma_1 \psi_1. \label{Eq3} $$

\section{Functional space and weak solution}
Let us formally introduce the operator
$$ A u = \gamma_1 (u, \psi_1)_{L_2(D)} \psi_1 +
\sum \limits_n \gamma_n (u, \psi_n)_{L_2(D)} \psi_n
+ \sum \limits_n \beta_n \coth(\beta_n a) (u, \chi_n)_{L_2(D)} \chi_n. $$
We write the equation (\ref{15}) in the form
\be (- i \gamma_1 - 1) (u, \psi_1)_{L_2(D)}\psi_1 - \beta_1 \cot(\beta_1 a) (u, \chi_1)_{L_2(D)} \chi_1 + A u = -2 i \gamma_1 \psi_1. \ee
Introduce the Hilbert space
$$ V = \left\{u \in L_2(D), \gamma_1 |(u, \psi_1)_{L_2(D)}|^2 +
\sum \limits_n \gamma_n |(u, \psi_n)|^2_{L_2(D)} \right.
+ $$
$$ \left. \sum \limits_n \beta_n \coth(\beta_n a) |(u, \chi_n)|^2_{L_2(D)} < \infty \right\} $$
with scalar product
$$ (u, v)_V = (u, v)_{L_2(D)} + \gamma_1 (u, \psi_1)_{L_2(D)}(v, \psi_1)^*_{L_2(D)} +
\sum \limits_n \gamma_n (u, \psi_n)_{L_2(D)}(v, \psi_n)^*_{L_2(D)}
+$$
$$ \sum \limits_n \beta_n \coth(\beta_n a) (u, \chi_n)_{L_2(D)}(v, \chi_n)^*_{L_2(D)}. $$
Then formulating the problem of finding a weak solution to equation (\ref{Eq3}) in the form
$$ (- i \gamma_1 - 1) (u, \psi_1)_{L_2(D)} (v, \psi_1)^*_{L_2(D)} - $$
$$ \beta_1 \cot(\beta_1 a) (u, \chi_1)_{L_2(D)} (v, \chi_1)^*_{L_2(D)} +
(u, v)_V = -2 i \gamma_1 (\psi_1, v)^*_{L_2(D)}, \forall v \in V, $$
applying the Riesz theorem we reduce the original problem in the space $ V $ to the equation

\be u - (i \gamma_1 + 1) (u, \psi_1)_{L_2(D)} A^{-1} \psi_1 - \beta_1 \cot(\beta_1 a) (u, \chi_1)_{ L_2(D)} A^{-1} \chi_1 = -2 i \gamma_1 A^{-1} \psi_1 \label{4} \ee Multiplying scalarly (\ref{4}) by $ \psi_1 $ and $ \chi_1 $ in $ L_2(D) $, we get \be (1 - (i \gamma_1 + 1) (A^{-1} \psi_1, \psi_1)_{L_2(D)} )(u, \psi_1) _{L_2(D)} + \beta_1 \cot(\beta_1 a) (A^{-1} \chi_1, \psi_1)_{L_2(D)} = \ee $$ -2 i \gamma_1 (A^{-1} \psi_1, \psi_1)_{L_2(D)}, $$ $$ -(i \gamma_1 + 1) (A^{-1} \psi_1, \chi_1)_{L_2(D)} (u, \psi_1)_{L_2(D)} + \beta_1 \cot(\beta_1 a) (A^{ -1} \chi_1, \chi_1)_{L_2(D)} (u, \chi_1)_{L_2(D)} = $$ $$ -2i \gamma_1 (A^{-1} \psi_1, \chi_1 )_{L_2(D)}. $$ Find from here $ (u, \psi_1)_{L_2(D)} $: $$ (u, \psi_1)_{L_2(D)} = $$ $$ ((1 - (i \gamma_1 + 1)(A^{-1} \psi_1, \psi_1)_{L_2(D)}) (1 + \beta_1 \cot(\beta_1 a)(A^{-1} \chi_1, \chi_1)_{L_2(D)}) - $$ $$ (i \gamma_1 + 1) \beta_1 \cot( \beta_1 a)(A^{-1} \psi_1, \chi_1)^2)^{-1} $$ $$ (-2 i \gamma_1(1 + \beta_1 \cot(\beta_1 a)(A ^{-1} \psi_1, \psi_1)_{L_2(D)})(A^{-1} \chi_1, \chi_1)_{L_2(D)} - $$ $$ \beta_1 \cot(\beta_1 a) (A^{-1} \psi_1, \chi_1)^2_{L_2(D)}). $$

\section{Resonance effect}

If for some value of $ k $ it holds that
\be (1 + (A^{-1} \psi_1, \psi_1)_{L_2(D)})(1 + \beta_1 \cot(\beta_1 a) (A ^{-1} \chi_1, \chi_1)_{L_2(D)} - \beta_1 \cot(\beta_1 a)
(A^{-1} \psi_1, \chi_1)^2 = 0, \label{5 } \ee
then
$$ (u, \psi_1)_{L_2(D)} = 2. $$
We will show that in the neighborhood of the value $ k = \frac{\pi}{a} $ the equation (\ref{5 }).
Let us write this equation in the form
$$ \frac{1}{(A^{-1} \chi_1, \chi_1)_{L_2(D)}} (1 + (A^{-1} \psi_1, \psi_1)_{L_2(D)}(A^{-1} \chi_1, \chi_1)_{L_2(D)}) + $$ $$ \beta_1 \cot (\beta_1 a) (1 + (A^{-1} \psi_1, \psi_1)_{L_2(D)} - \frac{(A^{-1} \psi_1, \chi_1)^2_{L_2( D)}}{(A^{-1} \chi_1, \chi_1)_{L_2(D)}}) $$ Considering that $$ (A^{-1} \psi_1, \chi_1)^2_{ L_2(D)} \leq (A^{-1} \psi_1, \psi_1)_{L_2(D)}(A^{-1} \psi_1, \psi_1)_{L_2(D)} $$
and
$$ (A^{-1} \psi_1, \psi_1)_{L_2(D)} \to 0, $$
$$ (A^{ -1} \chi_1, \chi_1)_{L_2(D)} \to 0 $$
as the diameter of the region $ D $ tends to zero,
and
$$ \cot(\beta_1 a) \to -\infty $$
as $ k \to \mu_1 - \frac{\pi^2}{a^2} $
we arrive at the fulfillment of equation (\ref{5 }) at some value of $ k $.

Thus, at this value of $ k $ the signal falling from $ -\infty $ in the cylinder $ Q_1 $ is practically not reflected and spreads in the direction $ -\infty $ in the cylinder $ Q_2. $

\vspace{1cm}

  \noindent {\bf Acknowledgments.}{The research is supported by the MSHE "Priority 2030" strategic academic leadership program}.

\end{document}